\documentclass[%
 reprint,
 amsmath,amssymb,
 aps,
pra,
]{revtex4-1}

\usepackage{graphicx}
\usepackage{dcolumn}
\usepackage{bm}
\usepackage{braket}
\usepackage[hidelinks]{hyperref} 
\usepackage{hyperref}
\def\euysoB*{$^{151}$Eu$^{3+}$:Y$_2$SiO$_5$}
\def\euyso*{Eu$^{3+}$:Y$_2$SiO$_5$}
\def\us*{$\mu$s}
\def\um*{$\mu$m}
\def\TAFC*{$1/\Delta$}
\def\etal*{\textit{et al.}}
\DeclareMathOperator{\sech}{sech}

\DeclareMathOperator{\sinc2}{sinc^2}    

\begin{document}

\preprint{APS/123-QED}

\title{Towards highly multimode optical quantum memory for quantum repeaters}

\author{Pierre Jobez$^{1}$}
\author{Nuala Timoney$^{1}$}
\author{Cyril Laplane$^{1}$}
\author{Jean Etesse$^{1}$}
\author{Alban Ferrier$^{2,3}$}
\author{Philippe Goldner$^{2}$}
\author{Nicolas Gisin$^{1}$}
\author{Mikael Afzelius$^{1}$}
\email{mikael.afzelius@unige.ch}

\address{$^{1}$Group of Applied Physics, University of Geneva, CH-1211 Geneva 4, Switzerland}
\address{$^{2}$PSL Research University, Chimie ParisTech, CNRS, Institut de Recherche de Chimie Paris, 75005 Paris, France}
\address{$^{3}$Sorbonne Universit\'{e}s, UPMC Univ Paris 06, Paris 75005, France}

\begin{abstract}
Long-distance quantum communication through optical fibers is currently limited to a few hundreds of kilometres due to fiber losses. Quantum repeaters could extend this limit to continental distances. Most approaches to quantum repeaters require highly multimode quantum memories in order to reach high communication rates. The atomic frequency comb memory scheme can in principle achieve high temporal multimode storage, without sacrificing memory efficiency. However, previous demonstrations have been hampered by the difficulty of creating high-resolution atomic combs, which reduces the efficiency for multimode storage. In this article we present a comb preparation method that allows one to increase the multimode capacity for a fixed memory bandwidth. We apply the method to a $^{151}$Eu$^{3+}$-doped Y$_2$SiO$_5$ crystal, in which we demonstrate storage of 100 modes for 51 $\mu$s using the AFC echo scheme (a delay-line memory), and storage of 50 modes for 0.541 ms using the AFC spin-wave memory (an on-demand memory). We also briefly discuss the ultimate multimode limit imposed by the optical decoherence rate, for a fixed memory bandwidth.
\end{abstract}

\maketitle

Quantum communication encompasses technologies that can be used to distribute quantum states, such as entangled photons, between remote users \cite{Gisin2007a}. Entangled states are resources for tasks such as quantum key distribution, quantum teleportation and distributed quantum computing. Long-distance quantum communication based on direct transmission of entangled photons is limited in distance due to unavoidable losses in the quantum channel \cite{Inagaki2013,Korzh2015,Gisin2015}. For instance, even state-of-the-art fibres have a non-negligible loss of 0.16 dB/km. Quantum repeaters provide a solution to the quantum channel loss problem, which is based on entanglement distribution between local nodes and entanglement swapping, at the cost of requiring more quantum resources \cite{Briegel1998}.

An attractive approach to quantum repeaters only uses linear optics for entanglement swapping and ensemble-based quantum memories as nodes. This approach was first proposed by Duan, Lukin, Cirac and Zoller (the DLCZ scheme) \cite{Duan2001}, and several improved schemes along the same lines have since then been studied (see \cite{Sangouard2011} and references therein). It was realized, however, that in order to reach minimal useful rates (e.g. 1 Hz or more) using DLCZ-derived schemes, some kind of multiplexing would be required \cite{Simon2007,Collins2007}. The entanglement distribution rate in these schemes scales linearly with the number of modes used for multiplexing. However, this requires quantum memories that can store large number of modes, which could be encoded in time \cite{Usmani2010}, frequency \cite{Sinclair2014} or space \cite{Lan2009}. All quantum memory schemes based on atomic ensembles can be used for efficient spatial multimode storage \cite{Grodecka-Grad2012}. In this article we focus on multiplexing in time, so-called temporal multimode storage, where different memory schemes do not perform equally in terms of efficiency. It has been shown that the number of temporal modes $N$ that can be stored scales differently with the available optical depth $d$ of the ensemble \cite{Nunn2008}. For instance, using stopped light based on electromagnetically induced transparency (EIT) the scaling is $N \sim \sqrt{d}$, while for the gradient echo memory (GEM) the scaling is $N \sim d$. The atomic frequency comb (AFC) memory we employ here has the best scaling, being independent of the optical depth, although it depends crucially on other parameters as will be discussed below.

The AFC memory is a quantum memory scheme based on the creation of an absorption profile that has a comb structure with periodicity $\Delta$ \cite{Afzelius2009a}. This can be done by frequency-selective optical pumping in inhomogeneously broadened ensembles, in particular in rare-earth-ion doped solids \cite{Tittel2010b, RiedmattenAfzeliusChapter2015}. The basic mechanism underlying the scheme is the AFC echo produced a time $1/\Delta$ after an incoming photon was absorbed by the comb. This AFC echo provides a delay-type quantum memory (no on-demand read out). To realize an on-demand memory one introduces a pair of control pulses in order to reversibly map the optical coherence onto a spin coherence, which is called an AFC spin-wave memory. Recently the first AFC spin-wave memories working at the single photon level with high signal-to-noise ratio were demonstrated \cite{Gundogan2015,Jobez2015,Laplane2015}, with storage times up to 1 ms \cite{Jobez2015,Laplane2015}.

In both AFC schemes the temporal multimode storage capacity is proportional to the number of peaks in the comb \cite{Afzelius2009a}. Often the total AFC bandwidth is fixed, in which case the number of peaks can only be increased by decreasing the periodicity $\Delta$, or in other words by making the basic AFC delay $1/\Delta$ as long as possible. Here we present an optical pumping sequence that allows one to create a high-resolution and close-to-ideal combs with many peaks, while minimizing the comb creation time. This allows us to significantly increase the multimode capacity, both for AFC echo and AFC spin-wave memory experiments in our Eu$^{3+}$:Y$_2$SiO$_5$ crystal.

For all DLCZ-type quantum repeaters it is essential to be able to choose the read-out time of the memory, so-called on-demand read out, otherwise the quantum repeater does not provide any speed-up as compared to direct transmission, hence they require the full AFC spin-wave memory. However, a recently proposed quantum repeater protocol is exclusively based on a AFC echo memory and feed-forward control \cite{Sinclair2014}. In terms of the AFC echo memory it requires a very long $1/\Delta$ delay time, since the delay has to be at least the same as the photon propagation time over the length of an elementary repeater link, in order for the feed-forward control to work. In Ref. \cite{Sinclair2014} optimal lengths of around 100 km were considered, requiring AFC delays of around \TAFC* = 500 \us* for a fibre-based link. Shorter links can be used, but might not be optimal for the scheme. Using our preparation method we achieve the longest $1/\Delta$ delays reported so far, reaching 51 \us* for 100 modes and an efficiency of 8.5\% in a 0.1 \% doped $^{151}$Eu$^{3+}$:Y$_2$SiO$_5$ crystal. We also briefly comment on the limit of efficient AFC echoes at high $1/\Delta$ delays in terms of the optical coherence time and the radiative lifetime. Finally, we experimentally demonstrate the spin-wave storage of 50 modes during 540 \us* with an efficiency of 1.6\%. 

This article is organized as follows. In section \ref{AFC} we present the AFC protocol theory for both the delay memory version and the on-demand spin-wave version. In section \ref{shaping}, we detail the new method that we developed to tailor an AFC with a high resolution. Finally, we show in section \ref{Results} the results obtained by using this new method.

\section{\label{AFC} Atomic frequency comb memories}

\subsection{\label{AFC_2L} AFC echo delay memory}

\begin{figure}[htbp]
\centering\includegraphics[width=8cm]{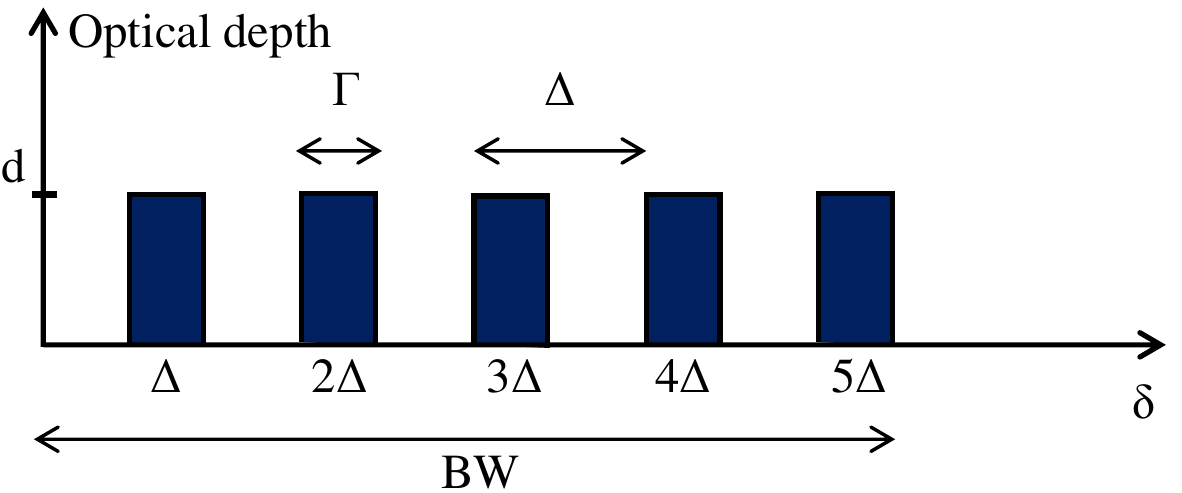}
\caption{Absorption spectrum of an AFC. Here $d$ is the optical depth of one peak, $\Gamma$ its width, $\Delta$ the periodicity and BW the bandwidth of the AFC.}
\label{peigne}
\end{figure}

The AFC memory is based on the creation of a periodic series of highly absorbing peaks. This comb-like structure can be created in a spectrally inhomogeneous ensemble by optically pumping away atoms at undesired frequencies, leaving absorption peaks with a periodicity $\Delta$ (see Fig. \ref{peigne}). The relative frequency $\delta_j$ of atom $j$ is given by $\delta_j=m_j \Delta$ in the limit of infinitely narrow peaks, where $m_j$ refers to the peak number (see Fig. \ref{peigne}) of atom $j$. A time $t$ after the photon was absorbed, the collective atom state is described by a Dicke-type large superposition state

\begin{equation}
\Ket{\Phi_{coll}} = \frac{1}{\sqrt{N_{at}}}\displaystyle{\sum_{j=1}^{N_{at}}}e^{-ikz_j+i2\pi\delta_{j}t}\ket{g_1...e_j...g_{N_{at}}},
\label{Dickestate}
\end{equation}

\noindent where $k=||\vec{k}||$ is the wave vector modulus of the incoming photon, $z_i$ the position of atom $i$ and $\Ket{g}$ ($\Ket{e}$) the ground (excited) state and $N_{at}$ the number of atoms. After a time $1/\Delta$, the inhomogeneous phase of atom $j$ is $2\pi m_j$, hence all terms in the superposition state are in phase. When all terms are in phase there is a strong collective coupling to the $\vec{k}$ light mode, which results in a strong emission probability in the direction given by the incoming photon (the AFC echo). The AFC echo only provides a delay-type memory, although tunable, since the comb periodicity cannot be re-programmed as the photon is stored. We note that all experiments demonstrating storage of quantum states of light in rare-earth doped crystals were realized using this method (see for instance Refs \cite{Saglamyurek2011, Clausen2011, Rielander2014,Bussieres2014, Tiranov2015a, Zong-QuanZhou2014}).

The AFC echo efficiency $\eta_{AFC}$ depends on the optical depth of the crystal $d=\alpha L$, where $\alpha$ is the absorption coefficient of the crystal and $L$ the length of the crystal. It also depends on the comb finesse, defined by $F=\Delta/\Gamma$ where $\Gamma$ is the width of one peak and $\Delta$ the comb periodicity (see Fig. \ref{peigne}). A higher comb finesse gives lower intrinsic inhomogeneous dephasing within one peak, but also a lower average optical depth $\tilde{d}=d/F$ \cite{Afzelius2009a}.  Additionally, it has been shown that the shape of each peak is very important to limit this intrinsic dephasing \cite{Sangouard2007}. The highest efficiency for a given optical depth $d$ is obtained for square peaks \cite{Bonarota2010}, where the intrinsic dephasing is given by $\sinc2(\pi/F)$. The echo efficiency can then be written as

\begin{equation}
\eta_{AFC} = \tilde{d}^2 e^{-\tilde{d}} \sinc2(\frac{\pi}{F}).
\label{AFC_EFF}
\end{equation}

For a given $d$, there is an optimal finesse $F^{opt}=\pi/\arctan(2\pi/d)$ which maximize the efficiency.
In this scenario the echo efficiency is bounded by 54\% due to re-absorption (the second term in the equation). This limit can be overcome by forcing the re-emission in the backward direction \cite{Sangouard2007,Afzelius2009a} or by placing the crystal inside an impedance-matched cavity \cite{Moiseev2010a,Afzelius2010a} to reach a theoretical limit of 100$\%$.\\

\subsection{\label{3LAFC} AFC spin-wave on-demand memory.}
\begin{figure}[htbp]
\centering\includegraphics[width=9cm]{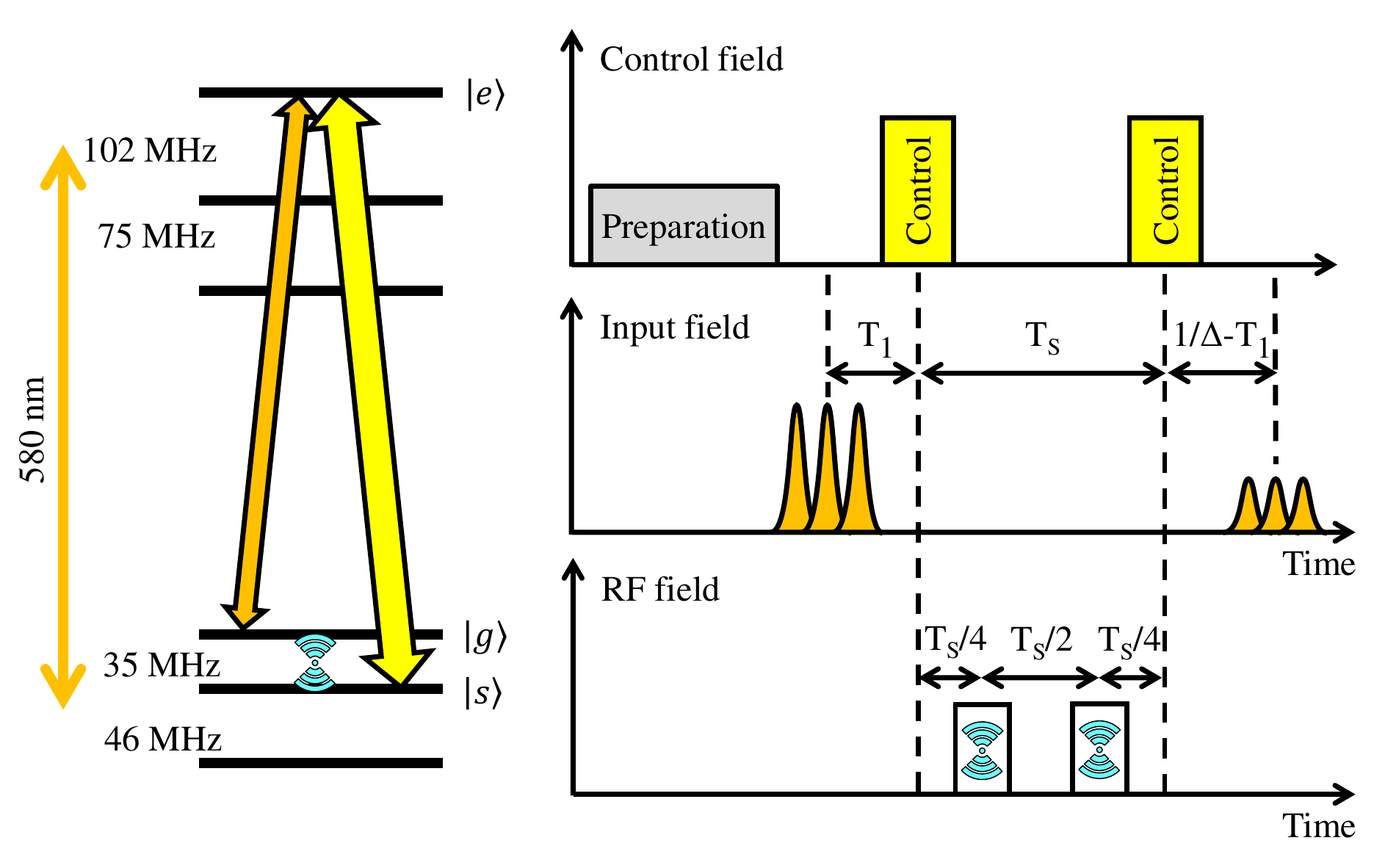}
\caption{Time sequence of the spin-wave AFC storage. The incoming input pulses are first mapped onto an optical collective excitation which is then transferred back and forth to a spin state using a pair of strong optical control pulses. A RF sequence can be applied during the spin-wave storage to increase the storage time beyond the limit otherwise given by the inhomogeneous spin broadening.}
\label{chronogramme}
\end{figure}

On-demand storage is realized by storing the coherence on a spin transition. This is achieved by transferring the population of the excited state to another ground state via a population inverting pulse (also called control pulse), see Fig.\ref{chronogramme}, which is applied before the AFC echo time $1/\Delta$. Application of a second control pulse, a time $T_S$ after the first one, re-establishes the optical coherence and leads to the emission of the output pulse after a total storage time $1/\Delta+T_S$. We call this an AFC spin-wave memory, which features on-demand read out. The first experimental demonstrations with spin-wave storage were made with classical pulses \cite{Afzelius2010,Timoney2012,Gundogan2013}, while the first steps towards storage of pulses at the single photon level were taken in Ref. \cite{Timoney2013}. Recently two experiments \cite{Gundogan2015,Jobez2015,Laplane2015} demonstrated a signal-to-noise ratio, at the signal photon level, that in principle would allow storage of quantum states of light.

The efficiency is generally lower than for the AFC echo at identical conditions. The spin transition is inhomogeneously broadened (of FWHM $\gamma_S$), which leads to a loss in efficiency $\eta_S$ due to inhomogeneous spin dephasing. Moreover, each control pulse usually has a non-unit transfer efficiency $\eta_T$. The AFC spin-wave  memory efficiency is then

\begin{equation}
\eta = \eta_{AFC} \eta_{T}^2 \eta_{S}=\eta_{AFC} \eta_{T}^2 e^{-\pi^2\gamma_S^2 T_S^2/(4\ln(2))},
\label{EFFsansRF}
\end{equation}

\noindent where the last factor holds for a Gaussian inhomogeneous spin linewidth.

As explained in the introduction the number of temporal modes of duration $\tau\sim \frac{1}{BW}$ that can be stored is limited by the number of peaks in the comb, which in turn depends on the periodicity $\Delta$ and the memory bandwidth (BW). But we also need to account for the duration of the control pulses $\tau_c$. The number of modes that can be stored then scales as

\begin{equation}
N_m \sim \frac{\frac{1}{\Delta}-\tau_c}{\tau}\sim BW(\frac{1}{\Delta}-\tau_c).
\label{Nmodemax}
\end{equation}

\noindent Realizing efficient and long AFC delays $1/\Delta$ is thus a key element for highly multimode AFC echo and AFC spin-wave storage.

\subsection{\label{RFsection} AFC spin-wave memory with spin control}

The limitation in spin storage time given by the inhomogeneous spin linewidth, see Eq. (\ref{EFFsansRF}), can be overcome with spin echo techniques \cite{Hahn1950}. The simplest sequence consists of two population-inverting radio-frequency (RF) pulses in resonance with the relevant spin transition frequency. More precisely, once the spin coherence is created, the atoms dephase with respect to each other due to their frequency disparity. After one RF population inverting pulse, the atoms begin to rephase. Only one pulse is necessary to rephase the atoms, but because this pulse is population inverting, we need to apply two of them in order to return to the initial  population situation. It is thus important to respect the pulse timing shown in Fig. \ref{chronogramme} to achieve a perfect rephasing. Using this RF sequence, the storage time is then limited by the spin coherence time $T_2^{spin}$

\begin{equation}
\eta=\eta_{AFC} \eta_{T}^2 e^{-2T_S/T_2^{spin}}.
\label{3L_EFF}
\end{equation}

We note that more complicated RF pulse sequences can be used. In a recent work \cite{Jobez2015,Laplane2015} we showed that a two-axis RF sequence, consisting of four RF pulses, was required in order to perform the AFC spin-wave memory, with spin control, when operating the memory at the single photon level. The two-axis sequence reduced spontaneous emission noise at the memory read out, which otherwise would have masked the weak output signal. Here we store classical pulses with many photons, in which case this simpler RF sequence is sufficient.

\section{\label{shaping} Shaping optimization of the AFC}

In the previous section we emphasized the importance of tailoring squarish AFC peak with an optimal finesse $F^{opt}$ to achieve efficient storage efficiency. To this end, our approach is to optically pump the atoms with a sequence of pulses that have a squarish frequency comb profile (the optimal shape) in the Fourier domain. This profile is necessarily the conjugate of the target absorption profile. A very good approximation of a squarish spectrum can be realized in practice by using adiabatic pulses \cite{Rippe2005,Seze2005,Minar2010}. These pulses are shaped with a secant hyperbolic function and frequency chirped with a tangent hyperbolic function. The amplitude $A(t)$ and the frequency $f(t)$ of the pulse can be written as:

\begin{equation}
A(t)=\sech(\beta t)\sin(\phi(t))
\label{sechyp_amp}
\end{equation}
\begin{equation}
f(t)=\frac{\dot{\phi}}{2 \pi}=f_0+\Delta_f \tanh(\beta t),
\label{sechyp_freq}
\end{equation}

\noindent where $\phi$ is the phase, $\Delta_f$ the chirp amplitude, $f_0$ the central frequency and $\beta$ a parameter related to the full width at half maximum (FWHM) of the pulse. 

To create an AFC profile with squarish peaks, one method is to burn each spectral hole one after the other by using sechyp pulses in series \cite{Bonarota2010}. The central frequency of each pulse is then detuned by $\Delta$ from the preceding pulse. In order to create the target AFC profile with a peak width of $\Gamma$ and a peak separation of $\Delta$, the spectral width of the pulse that will pump the absorption profile has to be $\Delta_f=\Delta-\Gamma$, the conjugate width of the target peak. We define $T_{prep}$ the time needed to generate one sequence (consisting of pulses in series), $\tau$ the single pulse length and $N$ the number of peaks. These two last parameters can be expressed as:

\begin{equation}
N=\frac{BW}{\Delta}
\label{NmodesBW}
\end{equation}
\begin{equation}
\frac{1}{\tau}=\frac{N}{T_{prep}}=\frac{BW}{T_{prep} \Delta}.
\label{tau_pulse}
\end{equation}

In order to tailor the square comb with a good resolution in the frequency domain, the single pulse Fourier limitation (or resolution) $1/\tau$ has to be significantly smaller than the target peak width $\Delta_f$. This limitation can be expressed roughly as follows:

\begin{equation}
\Gamma \gg \frac{1}{\tau}.
\label{limiting-condition}
\end{equation}

\noindent If the Fourier limitation $1/\tau$ becomes comparable or greater than the peak width, the AFC storage efficiency will be degraded. 

We here propose a modified technique. The idea is to create all the teeth in parallel (simultaneously) by using a sum of adiabatic pulses. This parallel sequence is thus constituted by a single pulse whatever the target AFC time $1/\Delta$. To be more precise, all the Fourier square components needed are generated in one single pulse with the following temporal profile:

\begin{equation}
A(t)=\sech(\beta t)\sum_{n=-N/2}^{N/2-1} \sin(\phi_n(t))
\label{parrallel_profile}
\end{equation}
\begin{equation}
f_n(t)=\frac{\dot{\phi_n}}{2 \pi}=f_0+n\Delta +\frac{\Delta_f }{2}\tanh(\beta t)
\label{parrallel_freq}
\end{equation}

\noindent which can also be written, after calculations:

\begin{equation}
A(t)=\sech(\beta t) \frac{\sin(N \pi\Delta t)}{\sin(\pi \Delta t)}\cos\Big[2\pi f_0 t+2\pi\frac{\Delta_f}{2\beta} \ln(\cosh(\beta t))\Big],
\label{sechyp_amp}
\end{equation}

\noindent where $f_0$ is the central frequency of the AFC. The resulting Fourier limit is then given by the sequence time $\tau=T_{seq}$. The Fourier resolution for the parallel method is then:

\begin{equation}
\frac{1}{\tau}=\frac{1}{T_{prep}}.
\label{parallel_derived_condition}
\end{equation}

\noindent Hence, the parallel method reaches the same Fourier limit $1/\tau$ as the serial method while using a $N$ times shorter preparation pulse (compare Eqs. (\ref{parallel_derived_condition}) and (\ref{tau_pulse})). In the experiments presented in this article, the number of teeth $N$ is larger than 100 for the longest $1/\Delta$ AFC echo times. The gain in memory preparation time is thus substantial.

\begin{figure}[htbp]
\centering\includegraphics[width=9cm]{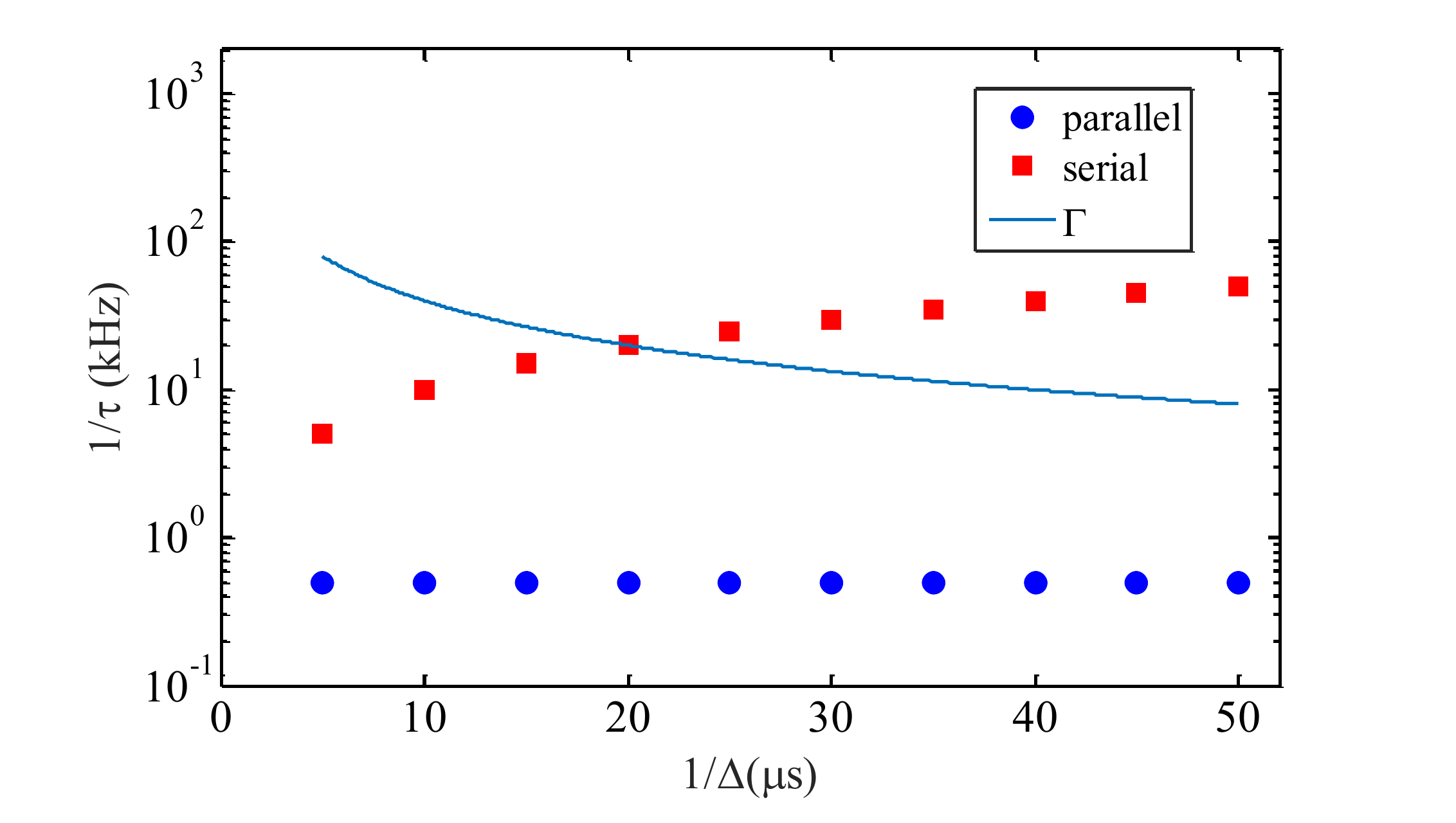}
\caption{Fourier limitation $1/\tau$ as a function of AFC echo time 1/$\Delta$, for the serial and parallel sequences discussed in section \ref{shaping}. The target AFC peak width $\Gamma$ is also plotted to illustrate the condition in Eq. (\ref{limiting-condition}).}
\label{method-comp}
\end{figure}

\begin{figure*}[htbp]
\centering\includegraphics[width=18cm]{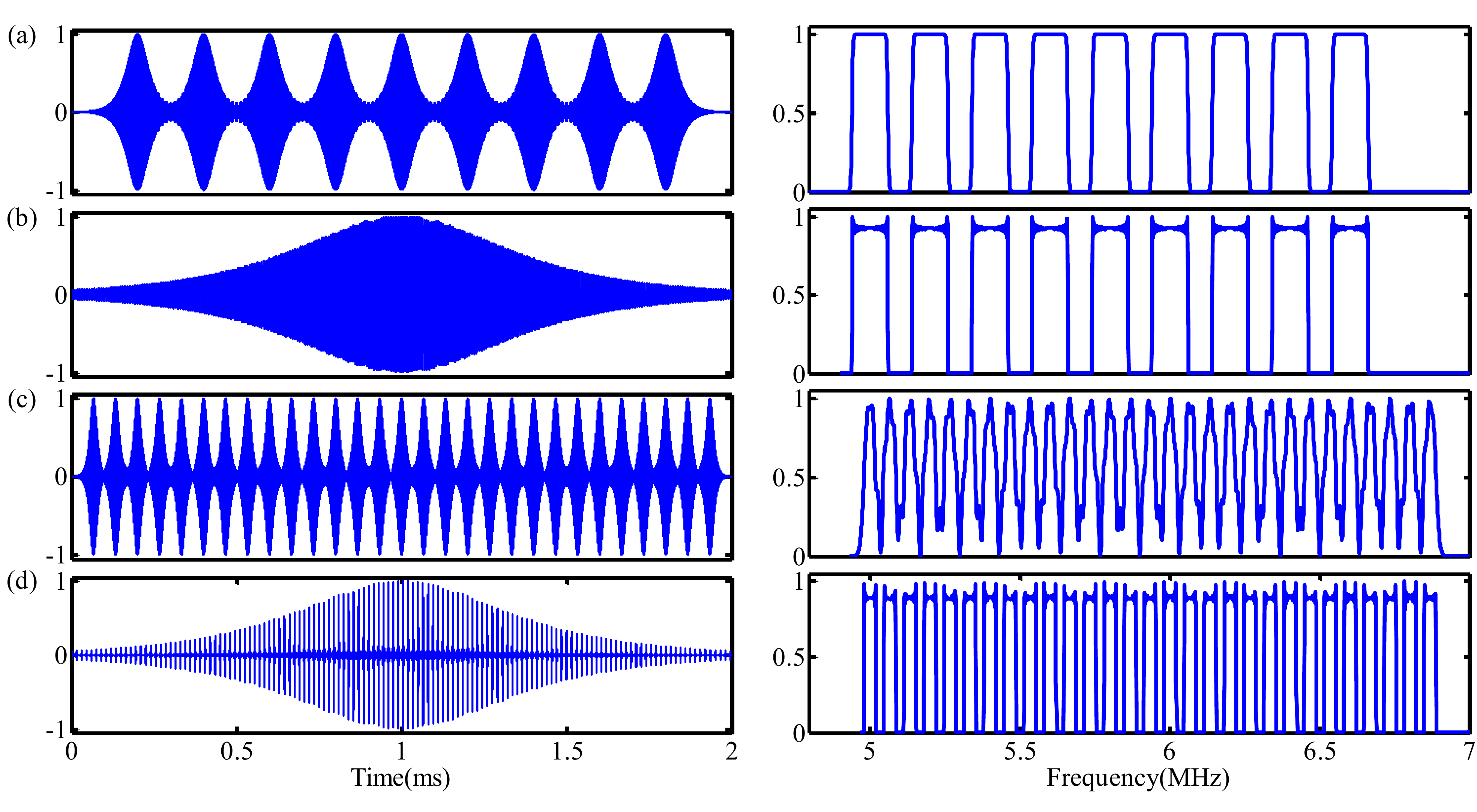}
\caption{We here compare the temporal profiles (left figures) and the corresponding Fourier spectra (right figures) for the serial (a and c) and parallel methods (b and d). Preparation sequences for two different AFC echo times are shown, $1/\Delta = 5 \mu s$ (a and b) and $1/\Delta = 15 \mu s$ (c and d). The vertical axes represent amplitudes in arbitrary units, both in time and frequency domain. Note that for the parallel method the top hat Fourier spectrum is slightly modulated due to the finite pulse width with respect to the total duration $T_{prep}$ = 2 ms.}
\label{Fourier-comp}
\end{figure*}

In Fig. \ref{method-comp}, the evolution of the Fourier limitation $1/\tau$ with the AFC echo time is represented for serial and parallel preparation sequence over the range of $1/\Delta$ values experimentally studied in this paper. The evolution of the peak width $\Gamma=\Delta$/F is also plotted as it has to be compared with the Fourier resolution (see Eq. (\ref{limiting-condition})) to ensure a good storage efficiency. The bandwidth and the finesse are both fixed to the experimental values BW = 2 MHz  and $F=2.5$ respectively. The preparation pulse duration is fixed to $T_{prep}$ = 2 ms. For the serial method, the Fourier limitation already reaches the target peak width $\Gamma$ in the range of 20 microseconds. At this crossing point the condition of inequality (\ref{limiting-condition}) is not fulfilled anymore. While for the parallel method, the Fourier resolution is still one order of magnitude smaller than $\Gamma$ up to 50 microseconds. This shows the capability of the parallel method to reach long AFC echo times while the serial method is quickly limited, for a given preparation pulse duration $T_{prep}$.

In Fig. \ref{Fourier-comp}, we compare the spectrum of the comb preparation pulse sequence for both the serial and parallel methods, using two different values of \TAFC*. For the serial method, we can clearly see that the spectrum gets degraded for \TAFC*=15 \us*, as compared to \TAFC*=5 \us*, while this is not the case for the parallel method. This degradation is explained by the single pulse duration reduction while increasing \TAFC*. Indeed, for the same bandwidth but with a longer \TAFC*, meaning a smaller spectral peak separation $\Delta$, the number of spectral holes to be burnt increases. As one pulse per spectral hole is needed for this method, the length of a single adiabatic pulse decreases. As a consequence, the short pulses needed for the comb of $1/\Delta=15\mu s$ lower the resolution of the resulting comb.

\section{\label{Results} Experimental results}

\subsection{\label{setup}Experimental setup}
The present AFC spin-wave memory is based on a custom-grown $^{151}$Eu$^{3+}$:Y$_2$SiO$_5$ crystal, with a $^{151}$Eu$^{3+}$ concentration of 0.1\%. We use the yellow $^7$F$_0$ $\rightarrow$ $^5$D$_0$ transition at 580.04 nm, which has an extremely narrow optical homogeneous broadening \cite{Koenz2003} at cryogenic temperatures. It also presents long spin coherence times \cite{Zhong2015,Arcangeli2014,Alexander2007}. The isotopically enriched $^{151}$Eu$^{3+}$ doping results in a larger optical depth as compared to a natural abundance of Eu$^{3+}$ isotopes \cite{Ferrier2016}. Here the optical inhomogeneous broadening of the $^7$F$_0\rightarrow ^5$D$_0$ transition is approximately 1.6 GHz and the overall absorption coefficient is $\alpha$=2.6 cm$^{-1}$. The inhomogeneous spin linewidth of the 34.5 MHz spin transition is Gaussian with a full-width at half-maximum of 27 kHz.

A schematic of the experimental setup is shown in Fig. \ref{setup}. The laser is an amplified and frequency-doubled diode laser at 1160 nm, producing 1.5 W at 580 nm, which is frequency stabilized to an optical cavity. This cavity is placed in a high vacuum chamber to limit acoustic and thermal fluctuations. This results in a laser linewidth of the order of 3 kHz for millisecond time scales. 

We apply the control and input pulses in two different spatial modes to filter out the strong control field for the detection. The input mode has a waist of 50 $\mu$m at the middle of the crystal. The control mode is applied with an angle of roughly five degrees with respect to the input mode and has a waist of 300 $\mu$m inside the crystal in order to achieve a good overlap with the input field. The control field coupling is also optimized to reach optical power of 600 mW inside the crystal.
An acousto-optic modulator (AOM) in double pass generates the necessary frequencies for the atomic preparation and the strong control pulses. This AOM is driven by an arbitrary function generator (AFG) during comb preparation and control pulse generation. As the AOM is mounted in double-pass, the light is modulated twice by the function sent from the AFG. This effect is compensated for by adapting the AFG output signal (see Appendix \ref{DblPassCorr}). A second AOM is used to generate the coherent input pulse.

\begin{figure}[htbp]
\centering\includegraphics[width=8cm]{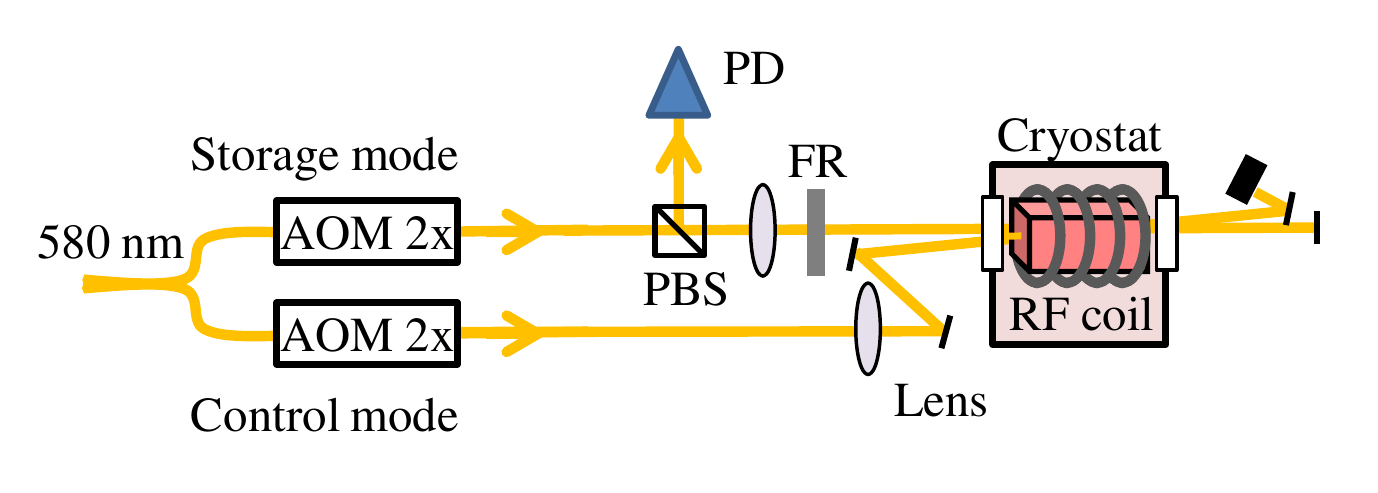}
\caption{Simplified sketch of the experimental setup for AFC memory experiments. The cavity-stabilized 580 nm light source is sent to two different AOMs: one generates the input pulses to be stored and one generates the pulses for the AFC preparation and the control field. The crystal is cooled to about 4 K in a cryostat. To increase the memory absorption, a double-pass configuration is realized through the crystal by using a polarizing beam splitter (PBS) and a Faraday rotator (FR). The RF coil placed around the crystal generates the 34.5 MHz field necessary for storage time extension through spin echoes.}
\label{setup}
\end{figure}

A 6-turn coil is placed around the crystal to generate the spin echo sequence discussed in section \ref{RFsection}. Finally, a magnetic field shield made with $\mu$-metal is placed around the cryostat chamber to reduce undesired static magnetic fields (the earth magnetic field for instance).

Before preparing the AFC, we first need to make a so-called class cleaning procedure (detailed in \cite{Jobez2014}) and then pump all the atoms into the $\ket{g}$ ground state (spin polarization sequence) by an optical pumping process. Through these two steps (of duration 0.5 s) the crystal is prepared on a 5 MHz bandwidth. Once this is done, the comb tailoring is ready to be realized. The class cleaning, spin polarization and comb preparation steps are all realized using the optical control mode.\\

\subsection{\label{AFC2L}Comb preparation and AFC echoes}
\label{2Lstorage}

We now experimentally compare the serial and parallel preparation sequences discussed in section \ref{shaping}. Each preparation sequence only pumps away a small fraction of the atoms in the targeted spectral regions, both for the serial and parallel methods. In principle one could pump deeper holes per sequence, but in order to reduce the effects of instantaneous spectral diffusion \cite{Koenz2003} we only excite a small fraction per pump cycle. To burn a high contrast AFC we then need to repeat each preparation sequence, in this particular case 100 times. The duration of the preparation sequence is fixed to $T_{prep}=$2 ms. After each preparation pulse, another 4 ms pulse is inserted on the $\ket{s}$-$\ket{e}$ transition in order to keep the spin storage state $\ket{s}$ empty of population. The input pulse is a Gaussian pulse with a FWHM of 900 ns. The input pulse contains a large number of photons (denoted a classical pulse), although its pulse area is much less than $\pi/2$. The bandwidth of the AFC is set to BW = 2 MHz, large enough as compared to the input pulse spectrum.

\begin{figure}[htbp]
\centering\includegraphics[width=8cm]{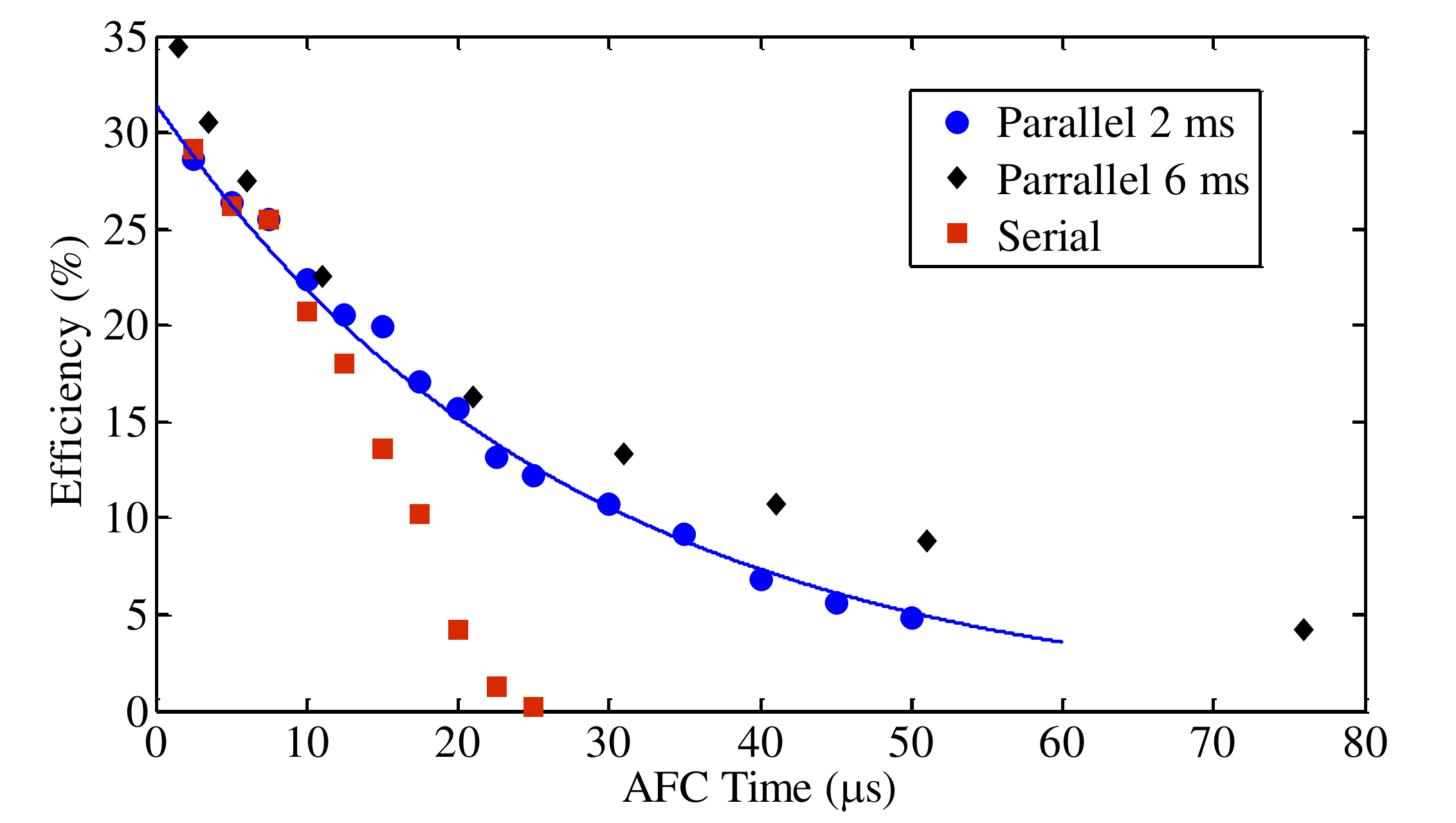}
\caption{AFC echo efficiency versus \TAFC* using the serial method (red squares) and parallel method  with $T_{prep}$ = 2 and 6 ms (blue circles and black diamonds). For the parallel method with $T_{prep}$ = 2 ms the efficiency is fitted to an exponential decay (blue line) with a time constant of 27.5 $\mu$s at $\exp(-1)$ of the maximum value. For the case $T_{prep}$ = 6 ms the decay is non-exponential.}
\label{Efficiencies}
\end{figure}

The resulting AFC echo efficiencies using both preparation methods are shown in Fig. \ref{Efficiencies}, with $T_{prep}$ = 2 ms. For short \TAFC* times the efficiency ($\simeq30\%$) is very close to the maximum theoretical value of 32$\%$ obtained for a squarish comb assuming the optical depth of $d$ = 4. This means that both preparation methods are able to create close to optimal combs for short values of \TAFC*. For longer \TAFC* echo times both methods produce AFC echoes with reduced efficiency, however, there is also a sharp drop for the serial method with respect to the parallel method in the \TAFC* = 10 - 20 $\mu$s region. This is consistent with the Fourier limit of the serial method with respect to the target teeth width, as shown in Fig. \ref{method-comp}. These data clearly show the advantage of the parallel method over the serial method for producing long AFC echo delays. We also recorded the AFC echo decay curve using the parallel method with $T_{prep}$ = 6 ms, also shown in Fig. \ref{Efficiencies}. The longer preparation pulse further increases the efficiency for delays or 30 $\mu$s of more, while it does not change the main loss of efficiency in the \TAFC* = 0 - 30 $\mu$s region. We must therefore consider alternative explanations for this loss.

As discussed in the introduction, long AFC echo delays are important for the multimode capacity of AFC-based memories. In addition, a recent quantum repeater protocol \cite{Sinclair2014} only relying on AFC echoes requires extremely long \TAFC* delays, of several hundreds of microseconds. The delays demonstrated here are, to our knowledge, the longest ones ever produced. A relevant question is whether our observed decay is limited by the optical homogeneous linewidth $\gamma_h$, and what is the ultimate limit set by the radiative lifetime of the transition? We first note that the smallest spectral hole that can be burnt into an inhomogeneous profile is $2\gamma_h$ \cite{Maniloff1995}, in the limit of low optical depth and low pump power. In this limit there would be a loss of AFC echo efficiency given by  

\begin{equation}
\eta_h = e^{-2\gamma_h/\Delta} = e^{-4/(T_2 \Delta)},
\end{equation}

\noindent where we used the definition of the optical coherence time $T_2$, $\gamma_h=2/T_2$. Note that here $\Delta$ is defined in Hz (cf. Eq. \ref{Dickestate}), while $\gamma_h$ is defined in rad/s. The coherence time fundamentally limits the maximum delays of the very efficient memories required for quantum repeater applications. Indeed, a 90\% efficient memory can only be obtained for delays shorter than $\leq$0.025$T_2$. If we take the ultimate limit where $T_2$ is limited by the radiative lifetime $T_1$, i.e. $T_2=2T_1$, then a 90\% memory can reach delays $\leq$0.05$T_1$. For our Eu$^{3+}$:Y$_2$SiO$_5$ crystal one could reach delays of about 100 $\mu$s ($T_1$ = 2 ms \cite{Koenz2003}) , while for a recently investigated Tm$^{3+}$:Y$_3$Ga$_5$O$_{12}$ crystal the limit would be 65 $\mu$s ($T_1$ = 1.3 ms) \cite{Thiel2014}.

Although our experiment demonstrates very long AFC echo delays, we are clearly far from the fundamental limit given by $T_1$. The decay obtained with $T_{prep}$ = 2 ms can be fitted to an exponential with a time constant that corresponds to an effective $T_2$ of 110$\mu$s, Fig. \ref{Efficiencies}. This is shorter than the optical coherence time of $\sim$400 \us* that we measured under our experimental conditions, using conventional photon echoes (extrapolated to low intensity in order to account for instantaneous spectral diffusion \cite{Koenz2003}). See also Ref. \cite{Ferrier2016} for a temperature study of the optical coherence time made on a sample from the same crystal boule. Our simple model above does not, however, take into account that we need to burn deep holes and the effect of power broadening. Several technical limitations could also explain the decay, such as laser frequency fluctuations and electronic noise in the AOM signals. Understanding the fundamental limitations in creating deep and precise comb structures is of great importance in order to increase the multimode capacity of AFC delay memories. Detailed theoretical and numerical calculations could shed more light on this problem.

\begin{figure}[htbp]
\centering\includegraphics[width=8cm]{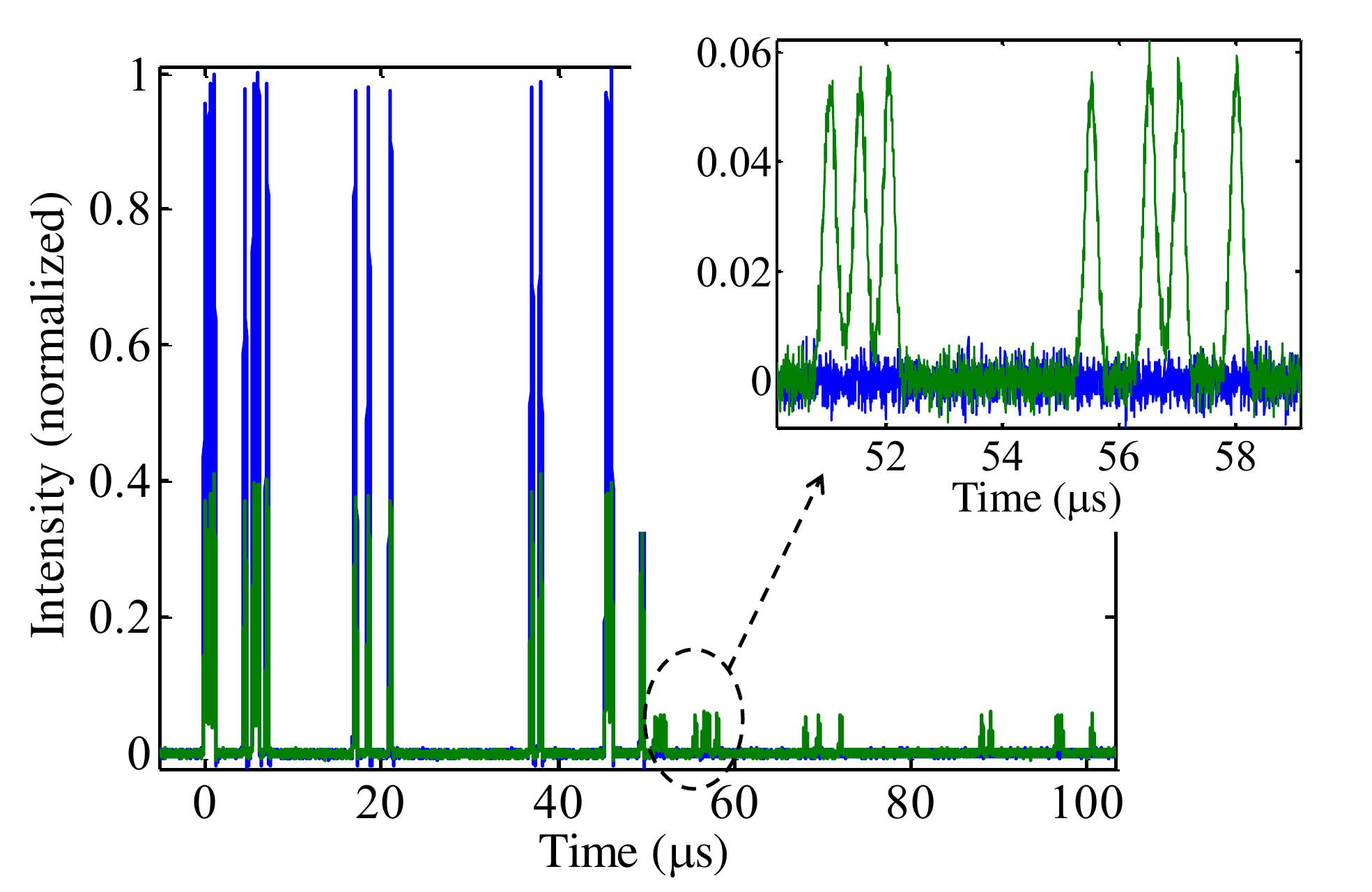}
\caption{AFC echoes of 100 temporal input modes for \TAFC*=51 \us*. The measured efficiency averaged over the temporal modes is 8.5$\pm$0.5\%.}
\label{multimode2L}
\end{figure}

To illustrate our current multimode capacity, when using the parallel method, we performed AFC echo experiments of 100 modes for 51 \us* (see Fig. \ref{multimode2L}). The preparation time was fixed to $T_{prep}$ = 6 ms and the preparation sequence repetition to 50, which in total does not change the total comb preparation time of 0.5 s. The AFC echo efficiency was then 8.5\%. The bandwidth of the AFC was set to 5 MHz, the largest value possible for this $\Lambda$ system of \euysoB* at zero external magnetic field. The larger BW allowed us to reduce the duration of each temporal mode to a FWHM of 300 ns. We randomly filled one over 8 temporal modes to avoid a high degree of excitation. Indeed, when storing a large number of classical pulses, a significant portion of the comb will be pumped into the excited state, thereby affecting the storage efficiency per mode. Note that this problem is only significant for classical storage, whereas for single photon level storage, this effect is negligible. 

\subsection{\label{AFC3L_results}AFC spin-wave memory}

Finally we show the high multimode capacity of the AFC spin-wave memory. To this end, we keep the same input and comb parameters given in section \ref{2Lstorage} for the multimode AFC echoes. We apply two adiabatic control pulses in the control mode (see Fig. \ref{setup}) with full power (600 mW), a 5 MHz chirp and a temporal duration of 14 \us*. These pulses were so-called HSH pulses \cite{Tian2011}, which have a more squarish amplitude profile than the adiabtic pulses used in for the AFC preparation (see Eqs. (5) and (6)). The HSH pulses then have a larger effective pulse area, which increased the transfer efficiency for these broadband pulses. During the spin wave storage, we apply the RF sequence discussed in the section \ref{RFsection} to achieve storage times in the millisecond regime. The inverting RF pulses are realized with 50 kHz chirped adiabatic pulses to transfer the 27 kHz inhomogeneous spin linewidth efficiently. 

\begin{figure}[htbp]
\centering\includegraphics[width=8cm]{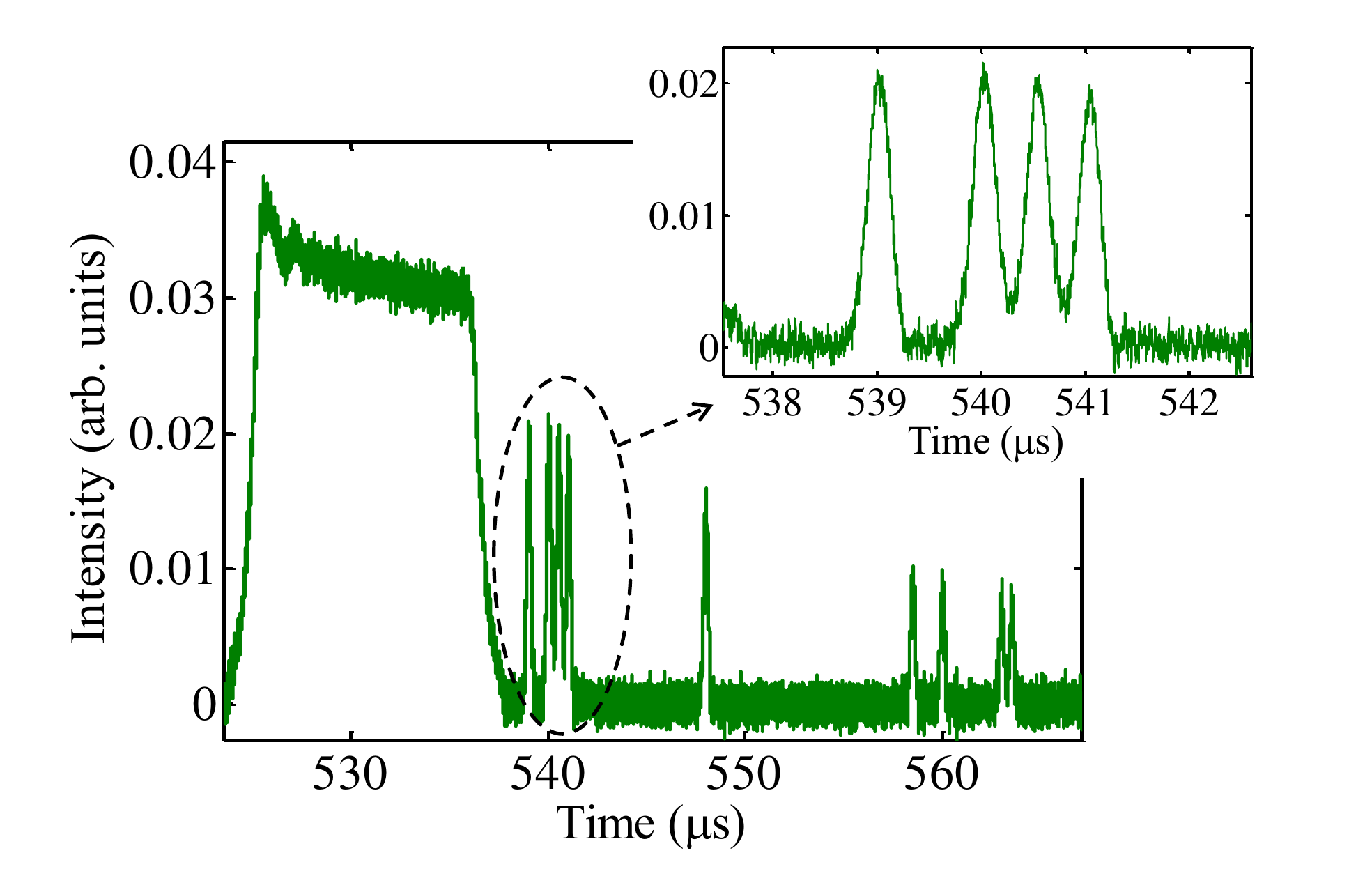}
\caption{Spin-wave storage of 50 temporal modes for 541 \us*. Here, \TAFC*=41 \us* and $T_S$=0.5 ms. The efficiency averaged over the temporal modes is 1.6 \%.}
\label{multimode3L}
\end{figure}

In Fig.\ref{multimode3L}, we show the spin-wave storage of 50 temporal modes for \TAFC*=41 \us* and $T_S$=500 $\mu$s. The average efficiency is 1.6$\pm$0.2 \% whereas the efficiency for the first 4 modes is about 2$\pm$0.2 \%. The control pulse transfer efficiency is about 55$\pm$0.5\% per pulse. In addition, we observed a spin coherence time of $T_2^{spin}$=3 ms. These limitations explain the efficiency obtained for the spin-wave storage efficiency as compared to the efficiency of 9$\pm$0.4\% for the AFC storage of 41 \us* (see Eq. (\ref{3L_EFF})). The decay of efficiency with the mode number stays unexplained for now and is still under investigation.

\section{\label{concl}Conclusion}

We presented an efficient method for creating precise, high-resolution atomic frequency combs with many comb teeth. Using this method we could demonstrate the longest AFC echo delays reported so far, up to 50 $\mu$s. We also applied this method to multimode storage, both using AFC echoes (100 modes stored) and a spin-wave AFC memory (50 modes stored) with a storage of 0.541 ms. We emphasize that although this work takes an important step towards highly multimode memories, the ultimate limit of the material is far from being reached. We therefore believe that the efficiency and multimode capacity can be significantly increased.

An interesting continuation of this work would be to extend these storage experiments to the quantum regime \cite{Gundogan2015,Jobez2015,Laplane2015}, particularly in the context of multimode DLCZ-type quantum nodes as proposed in Ref. \cite{Sekatski2011}. As entanglement generation rates increase linearly with the number of modes, this avenue is promising for efficient entanglement generation between remote quantum memories. 

\section{Acknowledgements}

We thank Walid Ch\'erifi for optical coherence time measurements and Nicolas Sangouard for theoretical support, as well as Claudio Barreiro for technical support. This work was financially supported by the Swiss National Centres of Competence in Research (NCCR) project Quantum Science Technology (QSIT), by the European Research Council (ERC-AG MEC) and by CIPRIS (People Programme (Marie Curie Actions) of the European Union Seventh Framework Programme FP7/2007-2013/ under REA Grant No. 287252). AF and PG also acknowledge financial support from DIM nano'K project RECTUS and Idex ANR- 10-IDEX-0001-02 PSL⋆. \\

\appendix
\section{AOM double pass compensation for arbitrary amplitude and phase modulation}
\label{DblPassCorr}

In this article we have considered a complex amplitude and phase modulation with multiple, simultaneous frequency components. This can in principle be achieved experimentally using an acousto-optic modulator (AOM) driven by a radio-frequency signal $g(t)$ corresponding to the signal that one wants to generate. In the linear regime of the response of the AOM, there is a linear relationship between first-order diffracted field amplitude $E_1(t)$ and the input field amplitude $E_0$,

\begin{equation}
E_1(t)=\kappa g(t) E_0,
\label{target_mod}
\end{equation}

\noindent where $\kappa$ is a coupling constant. But a single-pass AOM has a limited bandwidth due to the frequency-dependent beam deflection. The conventional solution to this problem is to operate the AOM in a double-pass configuration, which greatly extends its usable bandwidth. For a signal $g(t)$ that contains multiple frequencies this poses a problem, however, since the AOM can shift the light with any two different frequencies in the two passes. Technically the response function of the AOM is now $g^2(t)$. If we would feed the AOM with $N$ frequency signals with equal amplitudes we would have a power spectrum with a triangular amplitude spectrum, containing $2N$ frequencies, at the output of the double-pass AOM. We here present a method for performing an arbitrary amplitude and phase modulation, with multiple frequencies for instance, using an AOM in double-pass configuration. This method can be used, in particular, to create the comb generation pulse sequence of Eq. (\ref{sechyp_amp}).

If one wants to shape an arbitrary amplitude and phase modulation with an AOM of the form
\begin{equation}
f(t)=a(t)\cos\big[\omega_0t+\varphi(t)\big]
\label{target_mod}
\end{equation}
with a double-pass AOM, the electrical signal must be of the form
 \begin{equation}
g(t)=\sqrt{a(t)}\cos\Bigg[\frac{\omega_0}{2}t+\frac{\varphi(t)}{2}\Bigg].
\end{equation}
We can clearly see, though, that this method cannot be applied as soon as the amplitude modulation $a(t)$ takes negative values, as it is the case for the modulation described in Eq. (\ref{sechyp_amp}).

A method to circumvent this problem is to use the phase modulation as a tool to change the sign of the amplitude modulation, so that it always remains positive. For instance, if $\forall t, a(t)<0$, one sends the electronic signal:
\begin{equation}
g(t)=\sqrt{-a(t)}\cos\Bigg[\frac{\omega_0}{2}t+\frac{\varphi(t)}{2}+\frac{\pi}{2}\Bigg],
\end{equation}
so that in double-pass, the modulation is of the form
\begin{equation}
g^2(t)=-a(t)\cos\Bigg[{\omega_0}t+{\varphi(t)}+{\pi}\Bigg]=f(t).
\end{equation}
More generally, if we use the $sign(x)$ function, which is equal to 1 if $x>0$ and -1 if $x<0$, then sending the electronic signal
\begin{equation}
g(t)=\sqrt{|a(t)|}\cos\Bigg[\frac{\omega_0}{2}t+\frac{\varphi(t)}{2}+\frac{\pi}{2}\frac{1-sign\big[a(t)\big]}{2}\Bigg]
\end{equation}
will result in the proper modulation defined in Eq. (\ref{target_mod}) for a double-pass configuration AOM. 
\bibliography{Q:/Litterature/COMMONBIBFILE/qmcommon}
\bibliographystyle{apsrev4-1}

\end{document}